\documentclass[fleqn,usenatbib]{mnras}

\usepackage{newtxtext,newtxmath}

\usepackage[T1]{fontenc}

\DeclareRobustCommand{\VAN}[3]{#2}
\let\VANthebibliography\thebibliography
\def\thebibliography{\DeclareRobustCommand{\VAN}[3]{##3}\VANthebibliography}

\usepackage{graphicx}	
\usepackage{xspace}	    

\newcommand{\simgt}{\,\rlap{\lower 3.5 pt \hbox{$\mathchar \sim$}} \raise 1pt \hbox {$>$}\,}
\newcommand{\simlt}{\,\rlap{\lower 3.5 pt \hbox{$\mathchar \sim$}} \raise 1pt \hbox {$<$}\,}

\newcommand{\BE}{\begin{equation}}
\newcommand{\EE}{\end{equation}}
\newcommand{\BEA}{\begin{eqnarray}}
\newcommand{\EEA}{\end{eqnarray}}

\newcommand{\DV}{\ifmmode{\Delta v}\else $\Delta v$\xspace\fi}

\newcommand{\HI}{\ifmmode{\textsc{hi}}\else H\textsc{i}\fi\xspace}
\newcommand{\HII}{\ifmmode{\textsc{hii}}\else H\textsc{ii}\fi\xspace}

\newcommand{\Msun}{\ifmmode{M_\odot}\else $M_\odot$\xspace\fi}
\newcommand{\MUV}{\ifmmode{M_\textsc{uv}}\else $M_\textsc{uv}$\xspace\fi}
\newcommand{\fesc}{\ifmmode{f_\textrm{esc}}\else $f_\textrm{esc}$\xspace\fi}
\newcommand{\lya}{\ifmmode{\mathrm{Ly}\alpha}\else Ly$\alpha$\xspace\fi}

\newcommand{\nh}[1][]{\ifmmode{\overline{n}_\textsc{h}^{#1}}\else $\overline{n}_\textsc{h}$\xspace\fi}

\newcommand{\xHI}{\ifmmode{x_\HI}\else $x_\HI$\xspace\fi}
\newcommand{\xHImean}{\ifmmode{\overline{x}_\HI}\else $\overline{x}_\HI$\xspace\fi}
\newcommand{\xHIImean}{\ifmmode{\overline{x}_\HII}\else $\overline{x}_\HII$\xspace\fi}
\newcommand{\trec}{\ifmmode{t_\textrm{rec}}\else $t_\textrm{rec}$\xspace\fi}
\newcommand{\clump}[1][]{\ifmmode{C_\HII^{#1}}\else $C_\HII$\xspace\fi}

\newcommand{\Nion}{\ifmmode{\dot{N}_{\mathrm{ion}}}\else $\dot{N}_\mathrm{ion}$\xspace\fi}
\newcommand{\Rion}[1][]{\ifmmode{R_\mathrm{ion}^{#1}} \else $R_\mathrm{ion}$\xspace\fi}

\newcommand{\Tb}{\ifmmode{T_{21}}\else $T_{21}$\xspace\fi}
\newcommand{\aesc}{\ifmmode{\alpha_\mathrm{esc}}\else $\alpha_\mathrm{esc}$\xspace\fi}
\newcommand{\fescII}{\ifmmode{f_\mathrm{esc,10}^\textsc{ii}}\else $f_\mathrm{esc,10}^\textsc{ii}$\xspace\fi}
\newcommand{\fescIII}{\ifmmode{f_\mathrm{esc,7}^\textsc{ii}}\else $f_\mathrm{esc,7}^\textsc{iii}$\xspace\fi}
\newcommand{\astarII}{\ifmmode{\alpha_\star^\textsc{ii}}\else $\alpha_\star^\textsc{ii}$\xspace\fi}
\newcommand{\astarIII}{\ifmmode{\alpha_\star^\textsc{iii}}\else $\alpha_\star^\textsc{iii}$\xspace\fi}
\newcommand{\fstarII}{\ifmmode{f_{\star,10}^\textsc{ii}}\else $f_{\star,10}^\textsc{ii}$\xspace\fi}
\newcommand{\fstarIII}{\ifmmode{f_{\star,7}^\textsc{iii}}\else $f_{\star,7}^\textsc{iii}$\xspace\fi}
\newcommand{\tstar}{\ifmmode{t_\star}\else $t_\star$\xspace\fi}
\newcommand{\Mturn}{\ifmmode{M_\mathrm{turn}}\else $M_\mathrm{turn}$\xspace\fi}
\newcommand{\LX}{\ifmmode{L_X/{\dot{M}_\star}}\else $L_X/{\dot{M}_\star}$\xspace\fi}
\newcommand{\nuX}{\ifmmode{E_0}\else $E_0$\xspace\fi}
\newcommand{\AVCB}{\ifmmode{A_\mathrm{VCB}}\else $A_\mathrm{VCB}$\xspace\fi}
\newcommand{\ALW}{\ifmmode{A_\mathrm{LW}}\else $A_\mathrm{LW}$\xspace\fi}
\newcommand{\Mpcinv}{\ifmmode{\,\mathrm{Mpc}^{-1}}\else \,Mpc$^{-1}$\xspace\fi} 

\newcommand{\kp}{\ifmmode{k_\textrm{peak}}\else $k_\textrm{peak}$\xspace\fi}
\newcommand{\hp}{\ifmmode{h_\textrm{peak}}\else $h_\textrm{peak}$\xspace\fi}
\newcommand{\hMpc}{\ifmmode{\,h^{-1}\textrm{Mpc}}\else \,$h^{-1}$Mpc\xspace\fi}

\newcommand{\kms}{\,\ifmmode{\mathrm{km}\,\mathrm{s}^{-1}}\else km\,s${}^{-1}$\fi\xspace}
\newcommand{\cm}{\,\ifmmode{\mathrm{cm}}\else cm\fi\xspace}

\defcitealias{Mason2015}{M15}

\title[The brightest galaxies at Cosmic Dawn]{The brightest galaxies at Cosmic Dawn}

\author[C. A. Mason et al.]{Charlotte A. Mason$^{1,2}$\thanks{E-mail: charlotte.mason@nbi.ku.dk}, Michele Trenti$^{3,4}$ and Tommaso Treu$^{5}$
\\
$^{1}$Cosmic Dawn Center (DAWN)\\
$^{2}$Niels Bohr Institute, University of Copenhagen, Jagtvej 128, 2200 København N, Denmark\\
$^{3}$School of Physics, University of Melbourne, Parkville 3010, VIC, Australia\\
$^{4}$ARC Centre of Excellence for All Sky Astrophysics in 3 Dimensions (ASTRO 3D), Australia\\
$^5$ Department of Physics and Astronomy, University of California, Los Angeles, 430 Portola Plaza, Los Angeles, CA 90095, USA \\
}

\date{Accepted 2023 January 03. Received 2022 December 30; in original form 2022 July 30}

\pubyear{2015}

\begin{document}
\label{firstpage}
\pagerange{\pageref{firstpage}--\pageref{lastpage}}
\maketitle

\begin{abstract}
Recent JWST observations suggest an excess of $z\gtrsim10$ galaxy candidates above most theoretical models. Here, we explore how the interplay between halo formation timescales, star formation efficiency and dust attenuation affects the properties and number densities of galaxies observed in the early universe. To guide intuition, we calculate the theoretical upper limit on the UV luminosity function, assuming star formation is 100\% efficient and all gas in halos is converted into stars, and that galaxies are at the peak age for UV emission ($\sim10$ Myr).
This upper limit is $\sim4$ orders of magnitude greater than current observations, implying no formal tension with star formation in $\Lambda$CDM cosmology.
In a more realistic model, we use the distribution of halo formation timescales derived from extended Press-Schechter theory as a proxy for star formation rate (SFR). We predict that the galaxies observed so far at $z\gtrsim10$ are dominated by those with the fastest formation timescales, and thus most extreme SFRs and young ages. These galaxies can be upscattered by $\sim1.5$ mag compared to the median UV magnitude vs halo mass relation. This likely introduces a selection effect at high redshift whereby only the youngest ($\lesssim 10$\,Myr), most highly star forming galaxies (specific SFR$\simgt 30\,\mathrm{Gyr}^{-1}$) have been detected so far.
Furthermore, our modelling suggests that redshift evolution at the bright end of the UV luminosity function is substantially affected by the build-up of dust attenuation.
We predict that deeper JWST observations (reaching $m\sim30$) will reveal more typical galaxies with relatively older ages ($\sim100$\,Myr) and less extreme specific SFRs ($\sim 10\,\mathrm{Gyr}^{-1}$ for a $M_\mathrm{UV} \sim -20$ galaxy at $z\sim10$).
\end{abstract}




\section{Introduction}

Discovering when and how the first galaxies formed is still a major unsolved problem in modern astrophysics. Theoretical models predict the formation of the first sources, so-called, `Cosmic Dawn' was underway within the first few hundred million years of the universe's lifetime, $z\sim20-40$ \citep[e.g.,][]{Trenti2009,Bromm2011}.

The Hubble Space Telescope has allowed us to observe galaxies out to $z\sim10$, finding some that are relatively bright at rest-frame UV wavelengths \citep[e.g.,][]{Zheng2012,Coe2013,Oesch2016,Morishita2018,Bagley2022}. The rest-frame UV continuum is produced primarily by young ($\sim10$\,Myr), massive stars, and therefore the UV luminosity function (LF) traces recent star formation. Interpreting these observations requires comparison to theoretical modelling, and the UV LF has been modelled using a range of approaches, from `simple' but efficient semi-empirical models which can identify the main physical drivers of its evolution, \citep[e.g.,][]{Trenti2011,Tacchella2013,Mason2015,Sun2016,Yung2019,Mirocha2020}, to semi-analytic models which incorporate more physical processes and the realistic assembly of dark matter halos from merger trees \citep[e.g.,][]{Tacchella2018a,Yung2019,Hutter2021}, and finally hydrodynamical simulations which aim to model star formation and feedback processes self-consistently \citep[e.g.,][]{Ma2018,Vogelsberger2020}. Most $z\simlt8$ observations of the UV LF can be readily explained under the assumption of no redshift evolution in the star formation efficiency \citep[e.g.,][]{Trenti2011,Mason2015,Tacchella2018a,Bouwens2021}. However, relative to most theoretical models, an excess of bright galaxies is observed at $z\simgt8$ \citep{Bowler2020,Leethochawalit2022}.

The James Webb Space Telescope (JWST) is pushing our observational horizon even further, expanding our view to $5\mu$m \citep[with the NIRCam instrument,][]{Rieke2005}, making it possible to observe rest-frame UV emission from $z\simlt30$ galaxies, when the universe is just 100\,Myr old. Interestingly, recent works have shown JWST to preliminarily confirm HST galaxy counts at $z\sim 7-9$ \citep{Leethochawalit2022_JWST} but claimed discovery of a relatively high number of galaxy candidates at $z\sim10-17$ from early release science observations, suggesting the excess of sources relative to UV LF predictions extends to higher redshifts \citep{Castellano2022,Naidu2022,Adams2022,Morishita2022,Donnan2022,Atek2022,Finkelstein2022}. It is too early to draw definitive conclusions, as these sources are not spectroscopically confirmed and the photometric candidates are based on reductions using preliminary calibrations (indeed, some of the same sources are reported with significantly different fluxes by different groups), and cover small fields subject to cosmic variance \citep[e.g.,][]{Trenti2008}. Nevertheless, it is exciting to consider what the theoretical upper limit is on the UV LF at high redshift and how to interpret these early observations.

In the standard $\Lambda$ Cold Dark Matter ($\Lambda$CDM) paradigm, galaxies form as gas accretes into dark matter halos \citep[e.g. for recent reviews see,][]{Somerville2015,Wechsler2018}. However, the conversion of gas into stars appears to be inefficient, likely due to processes which heat and expel gas, generally described as `feedback' \citep[e.g.,][]{Silk1997}.
To investigate the maximum allowed UV LF as a function of redshift, we take a similar approach to \citet{Behroozi2018} who calculated a theoretical upper limit on stellar masses at $z>4$, in $\Lambda$CDM, assuming a 100\% star formation efficiency converting gas into stars. Due to order of magnitude uncertainties in measuring stellar mass, depending on assumed star formation histories \citep[e.g.,][]{Whitler2022}, and for easiest comparison to JWST data, we focus here on modelling the UV LF.
We will show that by assuming 100\% star formation efficiency and very young ages, this upper limit is at least 4 orders of magnitude above current observations and thus such early bright galaxies are not inconsistent with $\Lambda$CDM.

Motivated by recent work on the stochasticity of star formation, due to variation in halo formation histories, and its impact on the UV LF shape \citep[e.g.,][]{Ren2018,Ren2019,Mirocha2021,Furlanetto2022}, we assess whether UV bright galaxies can arise naturally as outliers from the general population, thanks to their very young ages and high star formation rates. We will show that current samples at $z\simgt10$, as a result of selection at relatively bright fluxes ($m_\mathrm{AB}<29$, $\MUV \simlt -19$), are dominated by extremely young, rapidly star-forming galaxies.

This paper is structured as follows: we describe our method for calculating the UV LF in Section~\ref{sec:methods}, we present our results in Section~\ref{sec:results} and conclude in Section~\ref{sec:conc}. We assume a flat $\Lambda$CDM cosmology with $\Omega_m=0.3,\,\Omega_\Lambda=0.7,\,h=0.7$ and magnitudes are in the AB system.

\section{Methods} \label{sec:methods}

\begin{figure}
\includegraphics[width=\columnwidth]{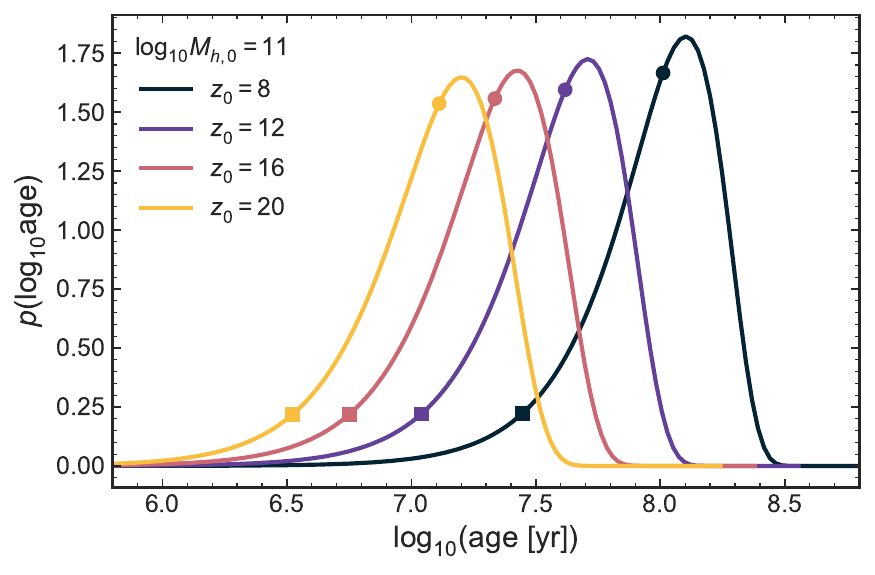}
\caption{Probability distribution of `ages' -- the time in which a halo with mass $M_{h,0}$ at $z_0$ assembled half its mass. The coloured lines show a range of observed redshifts for a halo of mass $10^{11}\Msun$. The circles show the position of the median formation time, the squares shows the lowest 10\% of ages.
\label{fig:p_age}
}
\end{figure}

UV continuum emission from galaxies is dominated by light from young stars ($\sim10$\,Myr), thus the UV emission traces the recent star formation history of a galaxy \citep[e.g.,][]{Stark2009}. High UV luminosities are produced by galaxies with very high star formation rates and young ages. The increasing presence of high equivalent width (EW) nebular emission lines with redshift implies galaxies in the early universe are very young and rapidly forming stars \citep[e.g.,][]{Labbe2013,Stark2013,Smit2014,Reddy2018,Boyett2022a,Stefanon2022}.

We model the UV LF following the prescription by \citet[][hereafter M15]{Mason2015}, which provides an accurate description of observations at $z\simlt8$. We assume that gas accretes at the same rate as dark matter and use extended Press-Schechter theory to calculate the evolution of dark matter halo properties \citep{Bond1991a}, as is commonly done by many other semi-empirical and semi-analytic models of the luminosity function \citep[e.g.,][]{Trenti2011,Tacchella2013,Behroozi2015a,Sun2016,Yung2019}.

A more realistic treatment of accretion rates can be obtained from merger trees in N-body simulations \citep[e.g.,][]{Liu2016a,Tacchella2018a,Hutter2021,Mirocha2021} or hydrodynamical simulations \citep[e.g.,][]{Vogelsberger2020,Wilkins2022} in order to model the high redshift UV LF. However, these simulations are volume-limited and $<$\,Gpc simulations do not provide sufficient statistics to understand how the rarest, brightest, sources form. For example, \citet{Mutch2016} identified only two galaxies as bright as GNz11 in the MERAXES semi-analytic model \citep[though, with increasing computing power, larger N-body simulations are becoming feasible for exploring rare high-redshift sources, e.g.,][]{Oogi2022}. In this work we use the distribution of halo formation times from extended Press-Schecter theory, assuming ellipsoidal collapse, to calculate timescales for star formation. Using this analytic approach we can build our model over an, in principle, unlimited range of halo masses.

To model the accretion rate of halos, we consider the "halo formation timescale" -- the time in which a halo doubles in mass. The distribution of halo formation times \citep[e.g.,][]{Giocoli2007} is:
\BE \label{eqn:pw}
p(w)dw = 2w \, \mathrm{erfc}\left(\frac{w}{\sqrt{2}}\right).
\EE
Here, $w$ is a time variable:
\BE \label{eqn:w}
w(M_h) = \sqrt{q}\frac{\delta_c(z_f) - \delta_c{z_0}}{\sqrt{\sigma^2(M_h/2) - \sigma^2(M_h)}}
\EE
where $\delta_c(z) \equiv \delta_c(0)/D(z)$ is the overdensity required for spherical collapse at $z$, where $\delta_c(0)\approx1.686$ and $D(z)$ is the linear growth factor, and $\sigma^2(M)$ is the variance of the linear fluctuation field, smoothed with a top-hit filter with scale $R=(3M_h/4\pi\overline{\rho})^{1/3}$ where $\overline{\rho}$ is the comoving density.

Figure~\ref{fig:p_age} shows the predicted formation time (which we will refer to as `age') distribution for a $10^{11}\Msun$ halo observed at $z\geq8$. Here we have converted from $p(w)dw$ to $p(\log_{10}t) d\log_{10}t$ numerically. A broad distribution of formation times is predicted for a given halo mass, with 10\% of halos having assembled $\sim5\times$ faster than the median of the distribution. As the majority of UV emission comes from $\sim10$\,Myr old stars, and SFRs are likely to track the halo formation timescale (assuming gas follows dark matter), the low age tail of this distribution is likely to be extremely UV bright. We note the similarity of this distribution to the age distribution derived from SED fitting of high redshift galaxies \citep[e.g.,][]{Whitler2022}, implying our assumption of star formation tracing halo mass accretion is reasonable.

We model galaxies' star formation histories (SFR) as increasing step functions as the halos accrete gas on the timescales defined above. This approximates rapidly rising SFRs seen in simulations at high redshift \citep[e.g.,][]{Finlator2011,Ceverino2018,Mirocha2021,Legrand2022}. We thus define the SFR as a function of the halo formation timescale. The SFR of a halo with mass $M_h$ (at lookback time $t_0$), which had mass $M_h/2^i$ at lookback time $t_i$, between timestep $t_i$ and $t_{i+1}$ is:
\BE \label{eqn:SFR}
\mathrm{SFR}(M_h, t_i, t_{i+1}) = f_b \frac{M_h}{2^i} \frac{\epsilon_\mathrm{SF}(M_h/2^i)}{t_{i+1}-t_i}
\EE
where $f_b \equiv \Omega_b/\Omega_m = 0.162$ is the cosmic baryon fraction and $\epsilon_\mathrm{SF}(M_h)$ is the star formation efficiency. The time interval $t_{i+1}-t_i$ is the halo formation (doubling) time, as a function of mass and redshift, as defined in Equations~\ref{eqn:pw} and~\ref{eqn:w}, and the factor $2^i$ accounts for the doubling of halo mass in each step. We refer the reader to Figure 2 by \citet{Mason2015} for a clear visualisation of these SFH models.

As in the \citetalias{Mason2015} model we assume $\epsilon_\mathrm{SF}$ is a function of halo mass but not of redshift \citep[consistent with other semi-analytic modelling efforts and with inferences from observations, see e.g.,][]{Trenti2011,Tacchella2018a,Bouwens2021,Harikane2022}. Due to the very young age of the universe at $z\simgt10$ and the rapid halo formation timescales that we will discuss below, this prescription for star formation histories leads to increasingly high SFR rates at early time, leading to SFH which are rapidly rising, mimicking the effects of recent and on-going star formation bursts.

To compute the UV luminosity of each galaxy we populate halos with stellar populations using the simple stellar population (SSP) models of \citet{Bruzual2003}. We assume a Salpeter inital mass function between $0.1-100\Msun$ and a constant stellar metallicity of $Z = 0.01 Z_\odot$. We neglect redshift evolution in metallicity under the assumption that very early star formation has already enriched the gas \citep[e.g.,][]{Trenti2009}, and note that metallicities do not significantly affect the non-ionizing UV continuum luminosity \citep[e.g.,][]{Schaerer2003}. The UV luminosity of a galaxy is given by:
\BE \label{eqn:LUV}
L_\textsc{uv}(M_h) = \sum_{i=0}^N \mathrm{SFR}(M_h, t_i, t_{i+1}) \int_{t_i}^{t_{i+1}} dt \; \mathcal{L}_\mathrm{ssp}(t)
\EE
where $\mathcal{L}_\mathrm{ssp}(t)$ is the luminosity at 1500\AA\ of a SSP of mass $1\Msun$ and age $t$. We consider two star formation steps, which \citetalias{Mason2015} demonstrated were sufficient to model the majority of the UV luminosity.

The stellar mass formed in a dark matter halo with mass $M_h$:
\BE \label{eqn:Mstell}
M_\star(M_h) = f_b M_h \sum_{i=0}^N \frac{\epsilon_{SF}(M_h/2^i)}{2^i}
\EE

Most semi-empirical models of the UV LF considered only a single star formation timescale \citep[e.g.,][]{Trenti2010,Mason2015,Park2019}. Here, we will sample timescales from Equation~\ref{eqn:pw} to calculate SFRs (Equation~\ref{eqn:SFR}). As discussed by \citet{Ren2019} and \citet{Mirocha2021}, adding scatter to the relation between UV luminosity and halo mass requires recalibrating $\epsilon_\mathrm{SF}$ to ensure models match the observed UV LFs. In this work we neglect this recalibration, as the effect only becomes significant for halos with masses $>10^{11}\Msun$, for which we expect very few at $z>10$. We thus use the $\epsilon_\mathrm{SF}(M_h)$ calibrated by \citetalias{Mason2015} at $z\sim5$ and do not expect this to have a strong impact on our results.

The UV luminosity function is then:
\BE \label{eqn:UVLF}
\Phi(\MUV) = \phi(M_h) \left|\frac{dM_h}{d\MUV}\right|
\EE
where $dM_h/d\MUV = \ln{10}M_h/2.5$ and we can invert the $L(M_h)$ relation in Equation~\ref{eqn:LUV} above to find $\MUV(M_h)$. In the following we assume the \citet{Reed2007} halo mass function (HMF) which was simulated specifically for $z\sim10-30$ halos, and is consistent with other HMFs \citep[e.g.,][]{Sheth2001} at lower redshifts. We use the python package \verb|hmf| \citep{Murray2013} to calculate the halo mass function and use the \citet{Eisenstein1998} matter transfer function.

\section{Results}
\label{sec:results}

\subsection{An upper limit on the UV LF}
\label{sec:results_maxLF}

To derive an upper limit on the UV LF we follow \citet{Behroozi2018} and consider the case of maximally efficient star formation, where all the gas is converted into stars $M_\star = f_b M_h$, i.e. the star formation efficiency is $\epsilon_\mathrm{SF}=1$, and ages for galaxies where they are at their peak emission of UV photons.

We calculate the maximum UV LFs at $z\sim8-20$ using Equations~\ref{eqn:LUV} and ~\ref{eqn:UVLF} above, assuming 100\% star formation efficiency and one period of constant star formation, $\mathrm{SFR} = f_b M_h \epsilon_\mathrm{SF}/ t_\mathrm{age}$, over $t_\mathrm{age}=10^7$ Myr. We assume $t_\mathrm{age}=10^7$ Myr as the majority of UV continuum photons are emitted during this period \citep[e.g.,][]{Schaerer2003}. This results in a specific star formation rate (sSFR) 100\,Gyr$^{-1}$, which is consistent with the highest sSFR at $z\simgt6$ \citep[][]{Stark2013,Santini2017,Endsley2021}.

\begin{figure}
\includegraphics[width=\columnwidth]{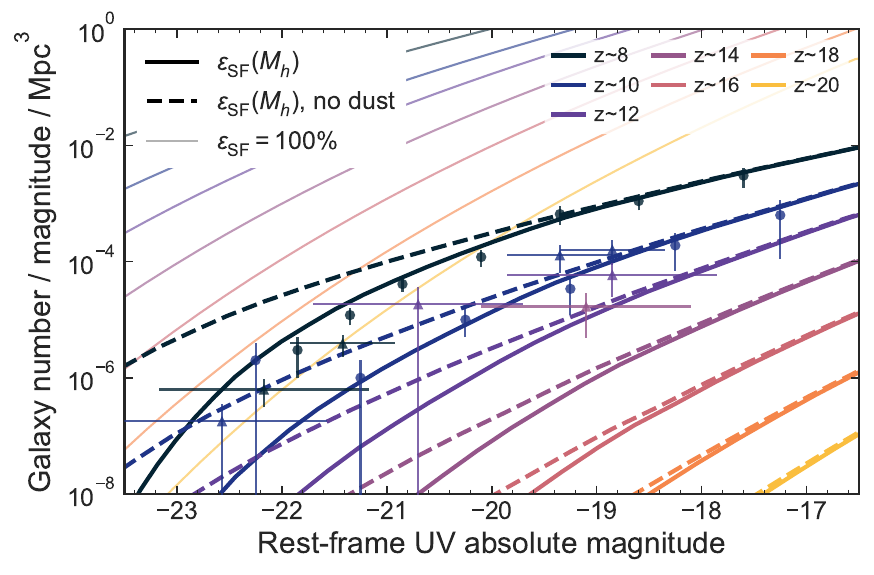}
\caption{UV luminosity function assuming 100\% star formation efficiency (thick solid lines) from $z\sim8-20$. For illustration, we also show a model which assumes a lower, halo mass-dependent, star formation efficiency \citepalias[extending the][model to $z>16$]{Mason2015}. Solid lines show the model including dust attenuation, dashed lines show the model without dust attenuation. Also shown (datapoints with errorbars) are recent constraints on the UV from HST, UltraVista and JWST at $z\sim8-10$ by \citet[][circles]{Bouwens2021} and $z\sim8-14$ by \citet[][triangles]{Donnan2022}.
\label{fig:UVLF}
}
\end{figure}

Our resulting LF is plotted in Figure~\ref{fig:UVLF}. We also plot the UV LF modelled as described in Section~\ref{sec:methods}, using a mass-dependent, but redshift-independent efficiency, $\epsilon_\mathrm{SF}(M_h)$, as derived by \citetalias{Mason2015}, with and without dust attenuation. The maximum theoretical LF is at least four orders of magnitude higher than the observations and the \citetalias{Mason2015} model, and the ratio between the maximum limit and the mass-dependent efficiency model increases with increasing redshift. This is because, in the \citetalias{Mason2015} model, galaxies have younger ages (because of higher accretion rates) and lower halo masses (because lower mass galaxies can be young and bright) at fixed \MUV and redshift. While the star formation efficiency does not evolve with halo mass in the model, the star formation efficiency as a function of \MUV is thus lower at fixed \MUV with increasing redshift.

We can also see dust attenuation is the dominant factor in the Schechter function shape of our model LF. In the \citetalias{Mason2015} model we use the \citet{Meurer1999} attenuation law $A_\mathrm{UV} = 4.43 + 1.99\beta$, modelling UV slope $\beta(\MUV,z)$ empirically from observations by \citet[][]{Bouwens2014a}. Past the limits of these observations we assume $\beta(\MUV,z>8)$ = $\beta(\MUV,z=8)$. However, in the early universe, it is plausible that limited dust has been produced in the majority of young galaxies. Simulations of very low metallicity $Z \sim 0.01Z_\odot$ galaxies predict limited dust formation by core collapse supernovae and UV slopes $\beta \sim -2.5$ \citep{Jaacks2018}. Early JWST observations have also found evidence for such blue UV slopes at $4<z<7$ \citep{Nanayakkara2022}. For $\beta\sim-2.5$ the attenuation is negligible, resulting in the dashed lines in Figure~\ref{fig:UVLF}, where the model follows the power-law shape of the HMF more closely. Thus a reduction in dust attenuation may play an important role in explaining excesses of bright $z\simgt8$ galaxies \citep[e.g.,][]{Bowler2020}. Indeed, we note that of the brightest $z\simgt11$ candidates found with JWST, all show very blue UV slopes $\beta \simlt -2.1$, indicating negligible dust attenuation \citep{Naidu2022,Atek2022,Donnan2022}. Though c.f. recent ALMA detections of dust continuum in `normal' $z\sim7$ galaxies \citep[e.g.,][]{Watson2015,Inami2022,Schouws2022}. Clearly, an improved understanding of dust attenuation is crucial for interpreting high redshift galaxy observations.

\subsection{The brightest galaxies are the youngest}
\label{sec:results_young}

\begin{figure*}
\includegraphics[width=1\columnwidth]{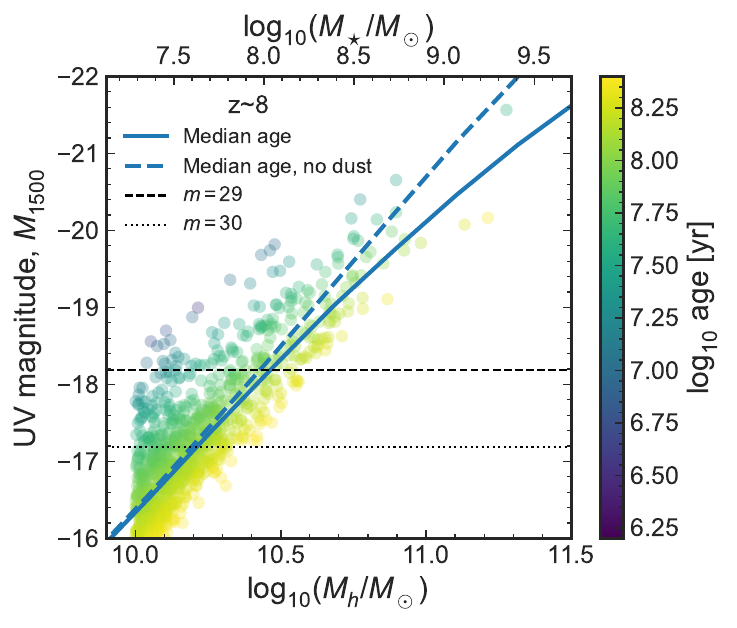}
\includegraphics[width=1\columnwidth]{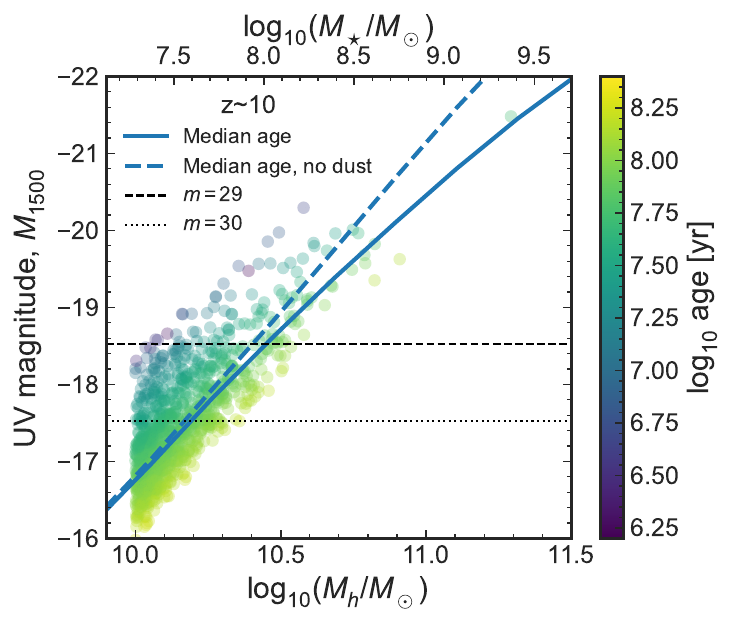}
\includegraphics[width=1\columnwidth]{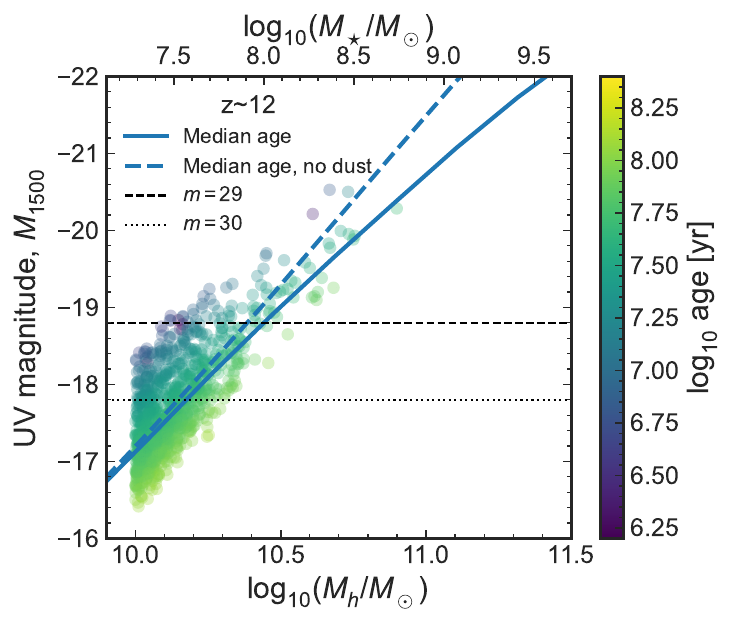}
\includegraphics[width=1\columnwidth]{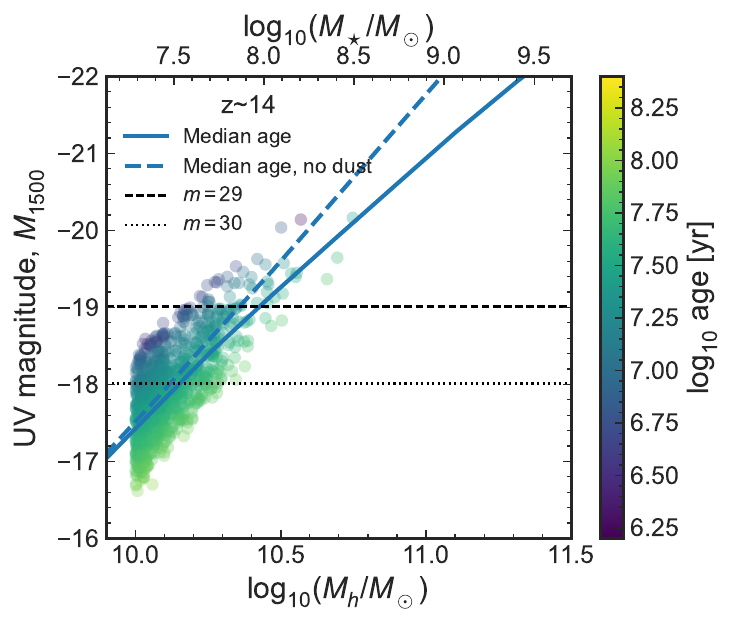}
\caption{UV magnitude, including dust attenuation, as a function of halo mass at $z
\sim8-14$. Points show 1000 sampled halos at each redshift, coloured by their formation time (time to form half their halo mass). We also show the relation obtained using the median formation timescale (solid blue line, including dust attenuation, dashed blue line, no dust attenuation). The dashed (dotted) horizontal line marks UV magnitude corresponding to $m_\mathrm{AB}=29$(30), typical limits for current and upcoming NIRCam imaging \citep[e.g.,][]{Merlin2022}. The majority of detectable galaxies lie significantly above the relation obtained using the median formation timescale -- young galaxies are upscattered to bright UV magnitudes.
\label{fig:Muv_Mh}
}
\end{figure*}

\begin{figure}
\includegraphics[width=\columnwidth]{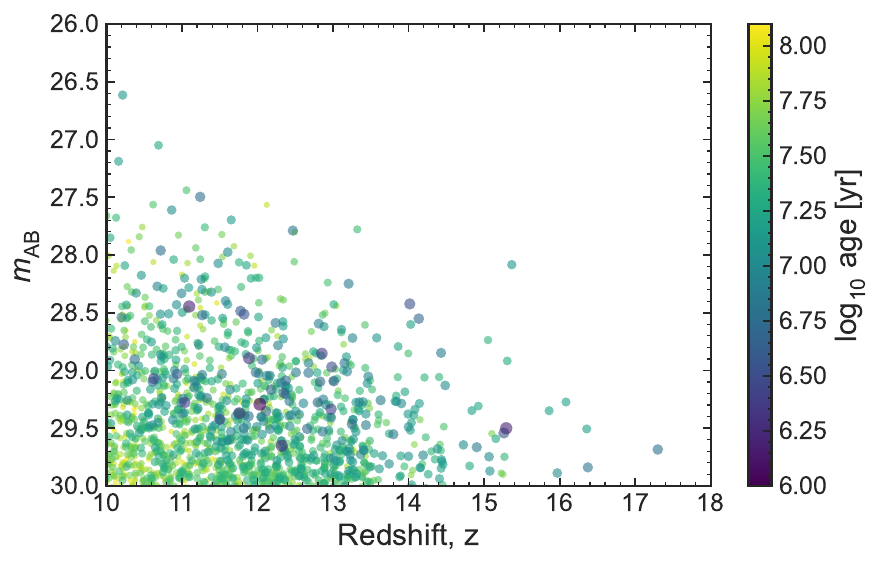}
\caption{Predicted galaxy apparent magnitudes (in the NIRCam band covering rest-frame 1500\AA) as a function of redshift, coloured by halo formation timescale, in a 1 sq degree survey. We add dust attenuation assuming the same $\beta(\MUV)$ relation as at $z\sim8$, if dust attenuation is negligible at high redshift galaxies in $>10^{10.5}\Msun$ halos may be $1-2$ mag brighter (see Figure~\ref{fig:Muv_Mh}).
\label{fig:survey}
}
\end{figure}

While the above upper limit is a useful illustrative constraint, 100\% star formation efficiency is very unlikely. Observational constraints on the stellar-to-halo mass relation and most theoretical models require $\epsilon_{SF} \times f_b \sim 1\%$ \citep[e.g.,][]{Behroozi2013a,Sun2016}. However, as described above, very young galaxies ($\sim10$\,Myr) will have maximally high UV continuum emission.

We now consider the hypothesis that the majority of galaxies observed at $z\simgt10$ are outliers in the populations, with very rapid star formation rates and young ages. This hypothesis has already been suggested as an explanation for the highest redshift spectroscopically confirmed galaxy, GNz11 \citep{Oesch2016,Mutch2016}. To test this hypothesis we now use the distribution of halo formation times (Equation~\ref{eqn:pw}) to calculate SFRs.

Figure~\ref{fig:Muv_Mh} shows halo mass versus UV magnitude at a range of redshifts. Here we sample 1000 halos with mass $M_h > 10^7 \Msun$ and draw halo formation timescales for each halo in order to calculate SFRs and UV luminosity as described in Section~\ref{sec:methods}. We compare these sampled galaxies to the $M_\mathrm{UV} - M_h$ relation obtained using the median halo mass distribution. We note that we expect UV luminosities to increase at fixed mass with increasing redshift, as halos assemble faster at higher redshift (see Figure~\ref{fig:p_age}).

We see the majority of galaxies that are observable within current limits are outliers in the \MUV-$M_h$ relation, as previously shown by \citet{Ren2019}. Here we have demonstrated that it is young ages which significant upscatter galaxies to UV magnitudes up to 1.5 mag above the median relation. Thus, current studies with JWST can probably only detect the youngest, highly star forming tip of the iceberg of the galaxy population \citep{Santini2022}. With deeper data throughout Cycle 1 and beyond, we may expect to reach $m_\mathrm{AB} \sim 30$. At these depths we predict it will be possible to observe older, more typical galaxies with ages $\sim100$\,Myr at $z\sim10-14$.

Finally, we make a forecast for the magnitude distribution of galaxies at $z>10$. In order to have enough statistics we forecast an area of 1 sq. degrees.
Based on this area we draw halos from the halo mass function in comoving volume bins corresponding to $\Delta z = 1$. For computational efficiency if the number of expected halos is greater than 1000, we set the sampled number to 1000. We then sample halo formation times for each halos using Equation~\ref{eqn:pw} and then calculate the observed UV luminosity of each galaxy, adding dust attenuation as described in Section~\ref{sec:results_maxLF}. Figure~\ref{fig:survey} shows the distribution of apparent magnitudes (in the NIRCam band covering rest-frame 1500\,\AA) for galaxies in this area as a function of observed redshift and halo formation timescale. We assigned redshifts in each bin by drawing from a uniform distribution $U\sim[z_0-0.5,z_0+0.5]$.

This figure again demonstrates that with increasing redshift, we predict observable galaxies to be increasingly young, with ages $\simlt 10$\,Myr. At $z\sim10$ our model predicts $\MUV \sim -20$ galaxies will have sSFR$\simgt30\,\mathrm{Gyr}^{-1}$ \citep[as seen in the brightest population already at $z\sim7$,][]{Endsley2021}. As described above, at fixed mass or \MUV we expect galaxies to have increasingly young ages at higher redshifts due to more rapid accretion. However, there is also a selection effect in a magnitude-limited survey: at fixed mass we will only be able to see the youngest galaxies, the median galaxy will be older and fainter. Therefore, current JWST studies are likely to be observing the most extreme sites of star formation in the early universe. We predict that future surveys reaching $m_\mathrm{AB}\sim30$ will instead identify a higher proportion of `older' galaxies ($\sim100$\,Myr) at $z\sim8-12$, where we predict $\MUV \sim -19$ galaxies will have sSFR$\sim10\,\mathrm{Gyr}^{-1}$.

We note that if dust attenuation decreases at $z>10$, all of our results can be shifted up by $\sim1$ mag.

\section{Conclusions} \label{sec:conc}

We have explored how the interplay between star formation efficiency, galaxy ages and dust attenuation affects the observability of $z\simgt10$ galaxies. Our conclusions are as follows:
\begin{enumerate}
    \item To guide intuition, we derive a theoretical upper limit on the UV LF assuming maximally efficient star formation and young ages which maximise UV emission (10\,Myr). Recent $z>8$ observations, including candidates discovered in JWST/NIRCam data, which show an excess of bright galaxies above theoretical models for the UV LF, are fully consistent with this theoretical upper limit on the LF, which is at least four orders of magnitude higher than the data. Therefore, there is no formal tension between current JWST observations at $z>10$ and formation of galaxies inside $\Lambda$CDM dark-matter halos.
    \item We find dust attenuation drives the Schechter function shape of our UV LF model at $z\simgt10$. If dust attenuation is negligible in galaxies' first few 100\,Myr, this could explain recent observations indicating a power-law LF at $z\simgt8$. A better understanding of dust attenuation at $z>8$ is thus crucial for interpreting high redshift observations.
    \item Using the distribution of halo formation times we find the majority of currently detectable galaxies ($m_\mathrm{AB} < 29$) lie significantly above the median $\MUV-M_h$ relation, and are so bright because they are young and rapidly assembling. This suggests galaxies currently observed at $z\simgt10$ are likely to be the most extreme tip of the iceberg in terms of star formation, but are unlikely to be representative of the overall galaxy population.
    \item We predict that within Cycle 1 of JWST, deeper surveys that will reach $m_\mathrm{AB}<30$ will detect older ($\sim100$\,Myr) galaxies, more typical of the population at fixed mass.
\end{enumerate}

We predict early JWST observations of $z\simgt10$ galaxies will be dominated by young galaxies with very high star formation rates, which are scattered up to 1.5 mag above the median $\MUV-M_h$ relation. While ages are challenging to measure at high redshift from SEDs \citep[e.g.,][]{Laporte2021,Tacchella2022,Whitler2022}, this is consistent with the increasing presence of strong nebular line emission in the brightest galaxy samples at $z\simgt7$ \citep[e.g.,][]{Endsley2021}, indicating very recent star formation and high sSFR. It is also interesting to compare with early JWST results which find similar rest-frame UV and optical morphologies and sizes at $z\simgt7$ \citep[][]{Chen2022,Treu2022_morph,Yang2022}, consistent with the hypothesis these galaxies will have only recently formed the bulk of their stars with very high sSFR.

JWST observations will enable us to understand the most extreme star-forming systems in the early universe. We predict the bias towards observing very young ages and high specific star formation rates means we will detect increasingly compact, clumpy galaxies \citep[e.g.,][]{Vanzella2022} with very high EW line emission. We expect more typical early galaxies, with more mature ages ($\sim100$\,Myr), will become visible at $z\sim10-14$ in surveys that can reach $m_\mathrm{AB}<30$, bringing us closer to completing the picture of galaxy build up at Cosmic Dawn.

\section*{Acknowledgements}
We thank Dan Stark for useful discussions. CAM acknowledges support by the VILLUM FONDEN under grant 37459. The Cosmic Dawn Center (DAWN) is funded by the Danish National Research Foundation under grant DNRF140. MT acknowledges support by the Australian Research Council Centre of Excellence for All Sky Astrophysics in 3 Dimensions (ASTRO 3D), through project number CE170100013. TT acknowledges suport from NASA through grant JWST-ERS-1324.

\section*{Data Availability}

Tabulated version of the UV LF predictions are available online at: \url{https://github.com/charlottenosam/UVLF_model}



\bibliographystyle{mnras}
\bibliography{library} 







\bsp	
\label{lastpage}
\end{document}